\documentstyle[12pt]{article}
\begin{document}
\begin{titlepage}
\renewcommand{\thefootnote}{\fnsymbol{footnote}}

\begin{flushright}
TPI-MINN-94/41-T\\
UMN-TH-1322-94\\
hep-ph/9412398\\
\end{flushright}
\vspace{.6cm}
\begin{center} \LARGE
{\bf A Closer Look at Perturbative Corrections in the $b \rightarrow
c$ Semileptonic Transitions}
\end{center}
\vspace*{.4cm}
\begin{center}
{\Large
M. Shifman $^a$ and N.G. Uraltsev $^{a,b}$}
\vspace*{.6cm}\\
{\normalsize
$^a$ {\it  Theoretical Physics Institute, Univ. of Minnesota,
Minneapolis, MN 55455}\\
$^b$ {\it Petersburg Nuclear Physics Institute,
Gatchina, St.Petersburg 188350, Russia}\footnote{Permanent
address}
\vspace*{.3cm}\\
e-mail address:\\
{\it shifman@vx.cis.umn.edu}
\vspace*{.95cm}}\\
{\Large{\bf Abstract}}
\end{center}

\vspace*{.5cm}
We comment on  two recent calculations of the second  order
perturbative corrections in the heavy
flavor semileptonic transitions  within the
Brodsky-Lepage-Mackenzie approach. It is pointed out that the
results do not show significant
enhancement either in the inclusive $b\rightarrow c$ decays or in
the exclusive amplitudes at zero recoil provided that the expansion
parameter is chosen in a way appropriate to the
kinematics at hand. The
values of the second-order coefficients inferred from the  BLM-type
calculations
appear to be of order unity in the both cases.  Thus, in
both cases no significant
uncertainty in
extracting $V_{cb}$ can be attributed to perturbative  effects. The
theoretical
accuracy is mostly determined by the existing uncertainty in
$1/m_c^2$ nonperturbative
corrections in the
exclusive $B\rightarrow D^*$  amplitude, and, to a lesser  extent, by
the uncertainty in the estimated value of
the kinetic-energy  matrix element $\mu_\pi^2$ in the case of
$\Gamma_{\rm
sl}(b\rightarrow c)$.
The theoretical accuracy of the inclusive  method of determination of
$V_{cb}$ seemingly
competes with and even exceeds the experimental accuracy.

\end{titlepage}
\addtocounter{footnote}{-1}

\newpage

In the last two years considerable  progress has been achieved
in the
theory of nonperturbative effects in the
heavy flavor decays. This theory has been applied to
many  problems from $B$ physics, including such practically
important application as precise determination of $V_{cb}$ from
experimental data. Further work in this direction is under way.

The current level of understanding of the nonperturbative effects
is such that, surprisingly, calculation of the `trivial' perturbative
corrections due to hard gluon exchanges becomes the bottle neck of
theoretical analysis. The perturbative corrections are
often suspected to limit the theoretical accuracy. In most of the
processes of
interest only one-loop
corrections are fully calculated at present. To match the level of
accuracy obtained in the nonperturbative sector one needs to know
two-loop ${\cal O}(\alpha_s^2 )$ terms.

 Complete calculation of these terms typically is a rather
difficult task. Sooner or later it has to be addressed
since without the complete two-loop results no analysis of the heavy
flavor decays can be considered as fully conclusive. At present such
calculations are not available,  and one has to settle for less.

It is known for a long time that $b$, the first coefficient in the
Gell-Mann-Low function,  is a large numerical parameter in QCD.  In
many
cases,
when both ${\cal O}(\alpha_s)$ and ${\cal O}(\alpha_s^2 )$
corrections are known, the set of
$(\alpha_s/\pi )^2$ terms is  dominated by those
graphs that are related to the running of $\alpha_s$.
\footnote{Let us parenthetically note that in the {\em Coulomb
gauge} these are merely the graphs with the polarization operator
insertions in the gluon line \cite{pole}.}
The sum of all these graphs is proportional to
$b(\alpha_s/\pi )^2$, while all other two-loop graphs
yield  $(\alpha_s/\pi )^2$ with a coefficient of order one. This
observation gives rise
to the so called BLM hypothesis \cite{blm} according to which a good
idea of the $(\alpha_s/\pi )^2$ terms can be obtained by calculating
this ``running $\alpha_s$" subset of the two-loop graphs while
ignoring all other
diagrams.

In two recent stimulating papers the BLM approach has been used
for  estimating the $\alpha_s^2$ corrections in the $b\rightarrow c$
transition. In Ref.~\cite{eta} (initiated by works \cite{pole,bb}) this
correction was
considered in the exclusive $b\rightarrow c$
amplitude at zero recoil and was found to be small. (By small we
mean that the coefficient in front of $(\alpha_s /\pi )^2$ is of order
unity.) A similar calculation was carried out in Ref.~\cite{wise}
for the inclusive
width of the $b\rightarrow c $ semileptonic decay with the
conclusion that the $\alpha_s^2$ terms are large. This result
was interpreted (see \cite{eta}) as a demonstration of an
uncontrollable
character of the
perturbative series in the inclusive decays which allegedly blocks
\cite{Nupd}
accurate determination of $V_{cb}$ within the inclusive method.

These two assertions above --  small two-loop
corrections in the exclusive case and large in the inclusive one --
seemingly clash with each other if one combines them with the facts that (i)
in the small velocity (SV) limit \cite{oneloop} the total inclusive
probability is
completely saturated by the exclusive transitions
$B \rightarrow D^*$ and $B \rightarrow D$; (ii) in the
$b\rightarrow c$ semileptonic decays we are actually rather close to
the SV limit --
in the essential part of the phase space the velocity of the
$D$ ($D^*$) is small in the $B$ rest frame. The proximity to the SV
limit is substantiated  by experimental data \cite{ron} according to
which
only 25 to
30\% of the total probability leaks into non-elastic channels (i.e.
those
other than $B\rightarrow D l\nu$ and   $B\rightarrow D^* l\nu$).
The theoretical clarification of this experimental fact goes beyond the
scope
of the present paper and is discussed elsewhere.

In this letter we argue that the paradox above is superficial.
Analyzing the exclusive and inclusive channels in parallel
and expressing the perturbative series in both cases in terms of one
and the same coupling constant (the one which is {\em physically
relevant}, see below) we eliminate the apparent qualitative contradiction
between the results for inclusive and exclusive transitions
and, moreover, demonstrate that the $\alpha_s^2$ terms
are small, so that they cannot be reliably captured  within the BLM
approach {\em per se}.
The present consideration, thus, gives further support to the
observation \cite{vcb} that the most accurate way of extracting
$V_{cb}$
available now is the analysis of the inclusive semileptonic rate where
the
theory of the nonperturbative effects is much more predictive than
in the exclusive decays.

To elucidate our point we need to do some preparatory work.
Let us start from the corrections in the exclusive $b\rightarrow c$
amplitudes at zero recoil. The one loop result for the axial form factor
$\eta_A$ at zero recoil is well known \cite{oneloop},
\begin{equation}
\eta_A = 1 +\frac{\alpha_s}{\pi}\left( \frac{m_b+m_c}{m_b-
m_c}\ln\frac{m_b}{m_c}
-\frac{8}{3}\right)\;\;;
\label{3}
\end{equation}
The analysis of Ref.
\cite{NU} establishes that the normalization point of the gauge
coupling $\alpha_s$ that is physically
adequate in this case is
\begin{equation}
m_0 =
\sqrt{m_cm_b}\;\;.
\label{mu0}
\end{equation}
 In other
words, if one expresses the perturbative series for $\eta_{V,A}$
in terms of $\alpha_s(m_0)$ no large numbers
appear in the expansion coefficients at order $\alpha_s^2$.
(Below, if not indicated explicitly, $\alpha_s$ means, by
definition, $\alpha_s(m_0)$.)
This leads to a numerical estimate \footnote{Some partial higher
order
calculations
of the $b\rightarrow c$ zero recoil amplitudes which existed in the
literature previously \cite{neupert,Nrev,Nupd} and gave rise to the
estimate
$\eta_A =0.986\pm 0.006$ were shown \cite{NU} to be irrelevant.}
$\eta_A\simeq 0.965$.

At the two-loop level
it was found in  Ref.~\cite{eta} that in the $\overline{\rm MS}$
scheme
\begin{equation}
\eta_A =
1-0.431\frac{\overline{\alpha}_s(m_0)}{\pi}-1.211
\left(\frac{\overline{\alpha}_s}{\pi}\right)^2 .
\label{4}
\end{equation}
The  values of the coefficients above refer to the
ratio of the quark masses
\begin{equation}
z\equiv m_c/m_b=0.3\;\; .
\label{5}
\end{equation}
We will adopt this value of $z$ throughout the paper  for the sake of
definiteness. All quantities referring to the $\overline{\rm MS}$
scheme are marked by bars. The following generic notation will be
used:
\begin{equation}
\eta_A = 1 + a_1\frac{\alpha_s}{\pi} + a_2\left(\frac{\alpha_s}{\pi}
\right)^2+\;...
\label{not}
\end{equation}

It is quite evident  that the  perturbative
coefficients starting from the second order depend on the scheme
used to define
the strong coupling. The number for  the two-loop coefficient  in
Eq.~(\ref{4})
is given  in the  $\overline{\rm MS}$
scheme often used in calculations performed in dimensional
regularization.
This scheme uses certain rather {\em ad hoc}
subtraction constants which are introduced for technical purposes
only, and is quite unphysical. Moreover, such a choice is definitely
unnatural for the calculations where one accounts only for  the
running of $\alpha_s$. The so-called $V$ scheme of BLM \cite{blm}
is physically preferable here
reflecting the real size of the effect. As a matter of fact,
the actual calculations within the BLM approach are routinely
performed just in this scheme (for details see
\cite{smvol,eta,wise}), and only at the final stage is the series
reexpressed in
terms of $\alpha_s$ in the  $\overline{\rm MS}$ scheme and
$\bar a_2$ introduced.
It is more adequate  to use the $V$-scheme strong coupling, at
least at the current stage when  the calculations beyond the BLM
level have not been done. In what follows we
will work with  the $V$-scheme $\alpha_s$,  unless otherwise
indicated; $\alpha_s$ no bar refers to the $V$ scheme while
$\overline{\alpha}_s$ is  reserved  for the
$\overline{\rm MS}$ coupling.

The one loop relation between $\overline{\alpha}_s$ and $\alpha_s$
is given by
\begin{equation}
\frac{\alpha_s(\mu)}{\pi} =\frac{\overline{\alpha}_s(\mu)}{\pi}
+\frac{5}{12} b \left(\frac{\alpha_s(\mu)}{\pi}\right)^2
\label{6}
\end{equation}
and, therefore, the  perturbative expansion in the
$\overline{\rm MS}$ scheme
$$
\eta_A =1+ \overline{a}_1 \frac{\overline{\alpha}_s(\mu)}{\pi}+
\overline{a}_2 \left(\frac{\overline{\alpha}_s(\mu)}{\pi}\right)^2
\;+\;...
$$
takes the form
\begin{equation}
\eta_A =1+\overline{a}_1 \frac{\alpha_s(\mu)}{\pi}+
\left(\overline{a}_2-\frac{5}{12}b\overline{a}_1\right)
\left(\frac{\alpha_s(\mu)}{\pi}\right)^2 \;+\;...\;\;
\end{equation}
so that $a_1 = \overline{a}_1$ and $a_2= \overline{a}_2-
\frac{5}{12}b\,\overline{a}_1\,$.
The first three terms in the expansion of $\eta_A$, Eq.~(\ref{4}), then
look as
follows for $m_c/m_b=0.3\,$:
\begin{equation}
\eta_A=1-0.431
 \frac{\alpha_s(m_0 )}{\pi} +0.286
\left(\frac{\alpha_s}{\pi}\right)^2
 +\;...\label{8}
\end{equation}
where $m_0$ is defined in Eq. (\ref{mu0}).
We see that the calculated second-order coefficient is in fact
noticeably less
than $1$, i.e. clearly
not enhanced. For this reason it cannot be taken even as an estimate
of the
second
order result because other  terms beyond the BLM approximation are
expected to be of
order 1.

 Now let us proceed to  the
inclusive semileptonic widths to bring the corresponding
analysis in line with the exclusive case.

Assume for a moment that  $(m_b-m_c)/(m_b+m_c)\ll 1$; then the
SV limit is obviously parametrically guaranteed, and   the
inclusive width is directly expressed in terms of $\eta_A$ and
$\eta_V$, the
zero recoil form factors of the vector and axial currents:
\begin{equation}
\Gamma_{\rm sl}=\frac{G_F^2(m_b-m_c)^5}{60\pi^3}
\left(\eta_V^2+3\eta_A^2\right)
|V_{cb}|^2\cdot \left[1+{\cal O}\left( \left(\frac{m_b-
m_c}{m_b+m_c}\right)^2
\right)\right]\;\;.
\label{9}
\end{equation}
This formula is nothing else than the text-book expression for the
neutron
$\beta$-decay width (with the electron mass neglected). Moreover,
in the SV
limit the deviation of the zero recoil vector form factor $\eta_V$
from unity
is also proportional to the same smallness parameter having the
meaning
of the
square of the typical velocity in the decay,
and thus can be neglected as well. Therefore in
the limit when $(m_b-m_c)/(m_b+m_c)\ll 1$ the perturbative
expansions for
the inclusive
width and for the  $B\rightarrow D^*\,\ell\nu$ zero recoil amplitude
squared  are
directly
related (essentially the same).

In the real world $m_c$ is significantly smaller than $m_b$.
However, as was mentioned above, in the semileptonic
$b\rightarrow c$ decays one effectively is not far
from the SV limit. The experimental evidence in favor of this fact is
the approximate saturation of the experimental inclusive  probability
by the two elastic channels.
It was pointed out in \cite{NU} that the coefficients in the
perturbative
expansion of $\eta_A$ do not deviate drastically from their values at
$m_c=m_b$; this observation was confirmed by explicit calculation in the BLM
approximation in Ref.~\cite{eta}.
Since near the SV limit the elastic channel saturation is
complete (this, of course, refers also to the perturbative corrections),
it is advantageous to write
the perturbative series for the inclusive width
in terms of the same coupling constant as
in the exclusive case, $\alpha_s(m_0)$.  This choice
rules out parametrically large two-loop coefficients
in $\eta$'s; by the same token it will kill such coefficients
in the inclusive width.

Let us examine the  result for the inclusive width \cite{wise} more carefully
taking advantage of the proximity to the SV limit.
Let us write down  the
perturbative expansion for the inclusive width in the form
$$
\Gamma_{\rm sl}=\frac{G_F^2 m_b^5}{192\pi^3}
|V_{cb}|^2 z_0(z) \kappa_\Gamma^2(z)\;\;,
$$
\begin{equation}
\kappa_\Gamma(z)=1+k_1(z)\frac{\alpha_s}{\pi}+
k_2(z)\left(\frac{\alpha_s}{\pi}\right)^2\;+\;...
\label{10}
\end{equation}
where $z_0(z)$ is the well known phase space factor.
Then one has
\begin{equation}
\kappa_\Gamma^2(z)=
\frac{1}{4}+\frac{3}{4}\eta_A^2(1) + {\cal O}\left((1-z)^2\right)\;\;.
\label{11}
\end{equation}

Comparing the expansions for $\eta_A$ and for $\kappa_\Gamma$
we get
$$
k_1(z)\,=\,\frac{3}{4}a_1(1)+{\cal O}\left((1-z)^2\right)\,=
\,-\frac{1}{2}+ {\cal O}\left((1-z)^2\right)\;\;,
$$
\begin{equation}
k_2(z)\,=\,\frac{3}{4}a_2(1)+ \frac{3}{32}a_1^2(1)+{\cal
O}\left((1-z)^2\right)\,=\,
\frac{3}{4}a_2(1)+\frac{1}{24}+{\cal O}\left((1-z)^2\right)\;\;\;.
\label{12}
\end{equation}
It is essential that terms ${\cal O}((1-z))$ are absent.
Examination of Eq. (\ref{12}) shows that there are no
reasons  to expect too big a  value of $k_2$ even
for the actual masses of $b$ and $c$. It is worth noting here that
Eq.~(\ref{11}) assumes the use of the ``pole" masses
for the  $b$ and $c$ quarks, and the corresponding definition of the
perturbative factor $\kappa_\Gamma^2$ in Eq.~(\ref{10}); otherwise
Eq.~(\ref{9}) does not hold.
The notion of the pole mass cannot be
consistently defined to all orders in the coupling constant, see
Refs.~\cite{pole,bb}. Perturbatively it can be  defined to a given
(finite) order, and
for our limited technical purposes of discussing the two-loop
perturbative corrections it is legitimate to use the corresponding
``two-loop pole" masses.

Of course, for Eq.~(\ref{12}) to hold the strong coupling must
be defined in the same way in both cases, inclusive and exclusive.
Taking into account the experimental proximity to the SV
limit in the semileptonic $b\rightarrow c$ decays, we will express
the perturbative
expansion for $\Gamma_{\rm sl}$ in terms of
$\alpha_s(m_0)$,
which will allow one a direct comparison with the exclusive decays.

Recently,  calculation of the $\alpha_s^2$ effects in
$\Gamma_{\rm
sl}(b\rightarrow c)$ associated with running of $\alpha_s$ has been
done
\cite{wise} based on a trick  suggested in \cite{smvol}. For
$m_c/m_b=0.3$ it was obtained that
\begin{equation}
\kappa_\Gamma^2=1-1.67\frac{\overline{\alpha_s}(m_b)}{\pi}-
15.1\left(\frac{\alpha_s}{\pi}\right)^2\;\;.
\label{14}
\end{equation}
According to Eq.~(\ref{6}) and the one-loop evolution of
$\alpha_s(\mu)$
this expansion in terms of $\alpha_s(m_0)$
(the proper choice for the kinematics at hand)
takes the form
\begin{equation}
\kappa_\Gamma^2=1-1.67\frac{\alpha_s(m_0)}{\pi}-
5.11\left(\frac{\alpha_s}{\pi}\right)^2\;\;+\;...\;\;\simeq \;1-0.16-
0.047+...
\label{15}
\end{equation}
where we used for numerical illustration the value
$\alpha_s(m_0)=0.3$
corresponding to the choice
$\overline{\alpha}_s(m_0)=0.24$
adopted in
Ref.~\cite{eta}. Perturbative corrections, thus, do not show
enhancement and, rather, seem to be well under control; no trace of
uncontrollable $10\%$ second-order corrections claimed in
\cite{Nupd} is seen.
We should note once again that the second-order coefficient in
Eq.~(\ref{15})
is not very large; it becomes even smaller if one considers
$\kappa_\Gamma$ instead of $\kappa_\Gamma^2$ which is
reasonable for the purpose of comparison with $\eta_A$ (see below).
If so, this result  cannot be taken too literally
because other
second-order corrections that are beyond BLM are expected to be of
a similar size.
\vspace*{.25cm}

Now we turn to the question of theoretical uncertainties in
determining $V_{cb}$.
Following the existing tradition to phrase the perturbative
corrections
to the
$B\rightarrow D^*\,\ell\nu$ rate in terms of corrections to the
amplitude
rather than to the rate itself, we also  write the perturbative
expansion for
the square root of the width, $\kappa_\Gamma= (\Gamma_{\rm
sl}^{\rm
pert}/\Gamma_{\rm sl}^0)^{1/2}$; this form is directly
related to extraction of $|V_{cb}|$:
\begin{eqnarray}
\eta_A=1-0.431
 \frac{\alpha_s(m_0)}{\pi}+ 0.286
 \left(\frac{\alpha_s}{\pi}\right)^2
 +\;...\;\;=\;1-0.041  +0.0026 +\;...
\nonumber \\
\kappa_\Gamma=1-0.835\frac{\alpha_s(m_0)}{\pi}-2.90\;\;
 \left(\frac{\alpha_s}{\pi}
\right)^2
 +\;...\;\;=\;1- 0.08 \;\, -0.026\;\;  +\;...
\label{17}
\end{eqnarray}
It is obvious that no significant theoretical uncertainty can be
attributed to
the perturbative corrections in both cases. Moreover, the genuine
relation between
the second-order coefficients can be established only after the
complete two-loop calculations are made: the BLM-type calculation
does not give a
large
contribution that would safely dominate the second-order result.

Continuing the analysis of the numerical aspect of the uncertainties it
is
appropriate to note that the nonperturbative corrections to the
inclusive
semileptonic widths are calculated \cite{buv,bs,prl} and give the
correction
\begin{equation}
\left(\delta\kappa_\Gamma\right)_{\rm nonpert} \simeq -2.5\%\;\;.
\label{18}
\end{equation}
This is smaller than the first order perturbative correction but
dominates the uncertainty in the $\alpha_s^2$ terms. A different
situation
occurs in  the zero recoil $B\rightarrow D^*$ amplitude. The
leading
nonperturbative corrections ${\cal O}(1/m_c^2)$ are not fully known
here (in a
model-independent way); a  reasonable estimate of these corrections
is about $-10\%$
\cite{vcb} with
the uncertainty of about $\pm 4\%$. Thus, they are definitely larger than
even the first order radiative corrections and, therefore, the exact value of
the second-order perturbative coefficient is not of much practical relevance
here since it is not abnormally large.

To complete our discussion of the purely perturbative uncertainties in
$b\rightarrow c$ transitions, let us mention that at present the value
of the
strong coupling at the scale $2.5\,{\rm GeV}$ is known, apparently,
with the
accuracy $\sim 10\%$. The corresponding uncertainty in
$\eta_A$ is thus $\sim 0.004$ and is definitely irrelevant in view of
large
nonperturbative uncertainties. The corresponding uncertainty in
$\kappa_\Gamma$
which enters determination of $|V_{cb}|$ from the inclusive width
constitutes about
$0.007$; being rather small and negligible at present, it may
contribute a
noticeable (through evidently not dominant) part in the overall
theoretical
uncertainty  in the near future. At the moment the lack of the
definite knowledge of the exact second-order coefficient also does
not
seem
to limit the theoretical accuracy of this most precise method
of
determination of  $|V_{cb}|$.

Two brief remarks concerning the nonperturbative corrections which
were mentioned in passing are in order here. The theoretical basis
for their calculation is provided by the Wilson operator product
expansion (OPE) \cite{wilson}. In calculating $\Gamma_{\rm sl}
(b\rightarrow c)$ OPE is used in the Minkowski domain which
assumes duality. Deviations from duality (they die off exponentially)
are controlled by the behavior (divergence) of the high-order terms
in OPE \cite{shif}; they can not be found from purely theoretical
considerations at present. On the basis of the phenomenological
information available it is reasonable to believe that these deviations
are negligible in the $b\rightarrow c$ transition (unlike in $c\rightarrow s$
decays \cite{dike}). They are not
included in the uncertainties in the nonperturbative terms which
were quoted above. Another aspect of OPE ignored so far in this
letter is the need for introducing a boundary point $\mu$
in any consistent OPE calculation with the subsequent separation of
the short-  and  large-distance contributions. The latter must be
subtracted from the coefficient functions calculated perturbatively
since they are accounted for in the matrix elements of
higher-dimension operators (condensate terms).
 In
the
calculations of the perturbative corrections to $\eta_A$
\cite{eta}
and, especially,    to the inclusive $c \rightarrow s\,\ell\nu$ width
\cite{wise0} the sizable part of the second-order coefficient comes
from
the low-momentum region of integration where perturbative
expressions for gluon
and quark propagators are not applicable. Therefore the numerical
estimates
derived from those calculations have no direct physical meaning: the
impact of
this region is completely accounted for  by nonperturbative
contributions and
must be excluded from the perturbative coefficients in the proper
treatment
(see \cite{pole,opt}). If it is done, the size of perturbative
corrections in  $c \rightarrow s\,\ell\nu$ decays reduces essentially.
In the case of $\eta_A$ this modification is numerically larger than
the whole second-order correction estimated in Ref.~\cite{eta}:
\begin{equation}
\eta_A(\mu)\simeq \eta_A+
\frac{\alpha_s(\mu)}{3\pi}\mu^2\left(\frac{1}{m_c^2}+
\frac{1}{m_b^2}+\frac{2}{3m_cm_b}\right)
\label{mudep}
\end{equation}
where $\mu$ is the infrared cutoff (for more detail see \cite{opt}).
The question of
the consistent simultaneous  account for both nonperturbative and
perturbative effects
\cite{opt} becomes important already at the level of the
second-order perturbative
corrections when the former are governed by the mass of $c$ quark,
due to the
existing numerical hierarchy among the corrections  in this case.
Although the need for such consistent
treatment based on the Wilson OPE is acknowledged, practically this
aspect is usually ignored in  HQET calculations (see, e.g.,
\cite{Nrev,NS}).
\vspace*{.3cm}

To summarize, we demonstrated that using the proper perturbative
expansion
parameter makes the second-order perturbative corrections equally
small both in
the $b\rightarrow c\,\ell\nu$ amplitude squared at zero recoil and in the
inclusive
$b \rightarrow c\,\ell\nu$ width. The existing BLM-type
calculations do not
show any particular enhancement of the second-order effects
associated with the
running of $\alpha_s$ if the scale is chosen in a way appropriate for
the SV
kinematics. For this reason the recent results \cite{eta,wise} on the
second-order perturbative
corrections in the inclusive and exclusive heavy quark transitions
can be
viewed mainly as the indication that these corrections are not large and
the
perturbative series are well controlled in both cases. More precise
results can
be obtained only by complete two-loop computations; they are
necessary  for the
ultimate improvement of the theoretical precision in the most
accurate method
based on the inclusive semileptonic width. For exclusive transitions
this
can not
improve the accuracy of extracting $V_{cb}$ because the uncertainty is
by far
dominated by rather
poorly known nonperturbative corrections. Still, the full two-loop
calculation
seems to be
desirable in this case as well: one can use the value of $V_{cb}$
obtained from the inclusive
widths
{\em to measure} the exclusive form factor with sufficiently high
accuracy and,
thus, to
obtain valuable information on the structure of nonperturbative
effects in the $b
\rightarrow c$ transitions in the situation when they are significant.

We emphasize that the concrete method of determining $V_{cb}$
from the inclusive width at present is based on the fact that the
expression for the
width valid in the SV kinematics works well for the actual quark
mass ratio -- the fact
known theoretically for a long time and now substantiated by
experimental studies
of the final charm states showing the relevance of the SV limit.
Therefore, the consistent analysis of corrections
requires using  the expansion parameters inherent to the SV
kinematics.

The result of recent calculations of potentially dominant
second-order
corrections \cite{wise} shows that the perturbative uncertainty
affecting the
extraction of $V_{cb}$  now does not exceed a percent level
and
is similar to the perturbative uncertainty in the exclusive zero recoil
transition, contrary to rather {\em ad hoc} claims existing in the
literature
\cite{Nupd,eta}. This uncertainty is smaller than the impact of the
nonperturbative corrections which are reliably calculated for
the inclusive $b\rightarrow c$
width.
On the contrary, the nonperturbative corrections seem to be much
larger in the
exclusive decays and, moreover, they carry a significant  uncertainty
that  can not be eliminated in a
model-independent
way at present. This limits the accuracy of the exclusive method of
extracting $V_{cb}$ but, also,
offers opportunities of learning  more about  details of the
nonperturbative
dynamics.

It seems justified to say  that with the recent progress in
the
theoretical treatment of the heavy flavor transitions the theoretical
accuracy in determination of $V_{cb}$ from $\Gamma_{\rm sl}(B)$
achieved  by
year 1995 exceeds the existing  experimental accuracy.

\vspace*{0.5cm}

{\bf ACKNOWLEDGMENTS:} \hspace{.4em} Useful discussions with I.
Bigi, B.~Smith,
A.~Vainshtein and M.~Voloshin are gratefully acknowledged.
This work was supported in part by DOE under the grant
number DE-FG02-94ER40823.

\end{document}